\documentclass[dvips]{article}

\usepackage{icrc2011}

\title{VHE Blazar Discoveries with VERITAS}

\shorttitle{W. Benbow:  VERITAS Blazar Discoveries}

\authors{Wystan Benbow$^{1}$ for the VERITAS Collaboration$^{2}$ }
\afiliations{$^1$Harvard-Smithsonian Center for Astrophysics, 60 Garden St., MS-20, Cambridge, MA, 02180, USA\\$^2$see J. Holder et al. (these proceedings) or http://veritas.sao.arizona.edu/conferences/authors?icrc2011}
\email{wbenbow@cfa.harvard.edu}

\abstract{The VERITAS array of 12-m atmospheric-Cherenkov telescopes
is used in an intensive observation program focused on the discovery 
of VHE ($E >$ 100 GeV) $\gamma$-ray-emitting blazars. 
Since VERITAS began full-scale operation in 2007, more than 1000 hours
of observation, on $\sim$90 targets, have been devoted to the
VHE blazar discovery effort and to timely follow-up in multi-wavelength (MWL)
observation campaigns.  These data have resulted in the discovery 
of VHE emission from 10 blazars (6
HBLs and 4 IBLs). A summary of these VHE  discoveries is presented.}
\keywords{VERITAS, AGN, Blazar, Gamma-ray, TeV, VHE}

\begin{document}
\maketitle


\section{Introduction}
\vspace{-0.2cm}
Blazars, a class of active galactic nucleus (AGN) 
with relativistic jets pointed along 
the line of sight to the observer, are the most numerous
class of identified VHE $\gamma$-ray sources.  Of the forty-one blazars
observed to emit VHE $\gamma$-rays, 73\% (30) are 
high-frequency-peaked BL\,Lac objects (HBL).  The remaining 11 blazars
were largely detected at VHE only during flaring episodes,
and include 5 intermediate-frequency-peaked 
BL Lac objects (IBLs), 3 low-frequency-peaked BL Lac objects (LBLs), and
3 flat-spectrum radio quasars (FSRQs).
Since VHE blazar studies probe the environment very near the central 
supermassive black hole of the AGN, such studies can
address a wide range of physical phenomena, including 
the accretion and jet-formation processes, and
can also be used to provide strong constraints on 
both the extragalactic background
light (EBL) and the intergalactic magnetic field.
VHE blazars have double-humped spectral energy distributions (SEDs).
The origin of the lower-energy peak is commonly explained as synchrotron 
emission from the relativistic electrons in the blazar jets.  
The origin of the higher-energy peak is
controversial, but is widely believed to be the result of 
leptonic processes, although hadronic scenarios are also
plausible. Given that blazar emission spans 
$\sim$20 orders of magnitude in energy and is highly
variable, MWL observations of these objects to measure 
both SED peaks contemporaneously are crucial for extracting the underlying
science.  Previous MWL studies have already yielded considerable
understanding of VHE blazars and their related science. 
It is particularly important to now develop these
implications further based on a large population of VHE blazars, 
from different blazar sub-classes, with contemporaneous SED modeling
in all cases.

\begin{table*}[t]
\begin{center}
\begin{tabular}{c | c | c | c |  c | c | c}
\hline
Blazar & $z$ &  Type & log$_{10}(\nu_{\rm synch})$       &  Flux [Crab \%] & $\Gamma$ & References \\
\hline
W\,Comae & 0.102 & IBL & 14.8 & Flares (9\% \& 23\%) &  $3.8\pm0.4$  & ATel \#1422, \#1565, \cite{WCom_paper1, WCom_paper2}\\
RGB\,J0710+591 & 0.125 & HBL & 21.1 & 3\% &  $2.7\pm0.3$ & ATel \#1941, \cite{RGBJ0710_paper}\\
1ES\,0806+524 & 0.138 & HBL & 16.6 & 2\% & $3.6\pm1.0$ & ATel \#1415, \cite{1ES0806_paper} \\
1ES\,1440+122 & 0.162 & IBL & 16.5 & 1\% & $3.4\pm0.7$ & ATel \#2786\\
RX\,J0648.7+1516 & 0.179 & HBL$^{*}$ & - & 2\% & $4.4\pm0.8$ & ATel \#2486 \\
RBS\,0413 & 0.190 & HBL & 17.0 & 2\% & $3.2\pm0.7$ & ATel \#2272\\
3C\,66A & ? & IBL & 15.6 & Varies (6\%) & $4.1\pm0.4$ & ATel \#1753, \cite{3C66A_paper1, 3C66A_paper2}\\
PKS\,1424+240 & ? & IBL & 15.7 & 5\% & $3.8\pm0.5$ & ATel \#2084, \cite{PKS1424_paper}\\
RGB\,J0521.8+2112 & ? & HBL$^{*}$ & - & Varies (5\%) & $3.5\pm0.2$ & ATel \#2260, \#2309\\
1ES\,0502+675 & ? & HBL & 19.2 & 6\% & $3.9\pm0.4$ & ATel \#2301\\
\hline
\end{tabular}
\caption{\footnotesize The blazars discovered at VHE by VERITAS. 
The blazar classification and synchrotron
peak frequencies are taken from \cite{Nieppola}, except for two cases (marked
with asterisks) where the classification is determined from 
VERITAS-led MWL studies.  The time-averaged VERITAS flux and 
photon index are also shown.}\label{discovery_table}
\end{center}
\end{table*}

\vspace{-0.2cm}
\section{The VERITAS Blazar Discovery Program}
\vspace{-0.2cm}
VERITAS began routine scientific observations with the full array in 
September 2007. The performance metrics of VERITAS 
include an energy resolution of $\sim$15\%, an angular resolution 
of $\sim$0.1$^{\circ}$, and a sensitivity yielding
a 5 standard deviation ($\sigma$) 
detection of an object with flux equal to
1\% Crab Nebula flux (1\% Crab) in $\sim$25 hours. 
For more details about VERITAS, see \cite{Holder_ICRC}.
VERITAS observations are performed for $\sim$1100 h 
each year, and a major goal of the VERITAS collaboration 
is to increase the number of identified VHE blazars.

Since the start of full-scale VERITAS operations,
observations of blazars have averaged $\sim$410 h per year.
During the first three years (2007-2010), the discovery of 
new VHE blazars was a primary focus of the VERITAS blazar program.  
Indeed, $\sim$80\% of the VERITAS blazar data 
came from discovery observations and follow-up observations
of any new sources.  In September 2010, VERITAS began to focus more
on deep studies of known VHE sources, and only 40\% of the blazar observations 
are now focused on VHE discovery.  Between September 2007 and June 2011,
a total of $\sim$1130 h of blazar-discovery data were taken.

The targets observed in the VERITAS blazar discovery program are largely HBLs,
but also include IBLs, LBLs and FSRQs.  The selection of the 
targets from 2007-09 is described in detail
in \cite{Benbow_ICRC09}.  The targets include EGRET-detected
blazars, together with
X-ray-bright HBLs, IBLs, and FSRQs recommended in
the literature (or objects first reported in later catalogs meeting the earlier
recommendation criteria), all subject to constraints based on
redshift ($z<0.5$), observability, and prior VHE exposure.
In February 2009, the Fermi-LAT collaboration
released its first catalog of 
MeV-GeV-bright blazars.  From this point onwards,
the VERITAS blazar discovery program has focused on
objects detected by Fermi-LAT, where either the power-law
extrapolation of the LAT photon spectrum (including
EBL absorption) indicates a likely VHE detection, or
where clusters of $>$10 GeV photons are found nearby.

In total, 20 VHE blazars are detected with VERITAS, including
10 discoveries.  The 10 blazars detected for the first
time at VHE with VERITAS are shown in Table~\ref{discovery_table}.
Following the discovery of a new VHE blazar, target-of-opportunity
(ToO) observations are initiated with VERITAS to
enable a better measurement of the VHE spectrum and light curve, as
well as with X-ray satellites (typically Swift) and optical/radio
observatories.  The goal of these MWL data is to provide
a contemporaneous spectral energy distribution (SED)
for modeling of the broadband emission.  Indeed,
such an SED has been generated for every VERITAS
VHE blazar discovery.

\vspace{-0.3cm}
\section{Pre-ICRC-2009 Discoveries}
\vspace{-0.2cm}
As of the 2009 ICRC, 11 VHE blazar detections
were reported by VERITAS.  Of these, 5 were VHE discoveries,
including the first 3 VHE IBLs.

{\bf 1ES\,0806+524} is an HBL at a redshift of $z = 0.138$.  It
was recommended as a likely VHE emitter by \cite{Costamante},
and was observed by VERITAS for 65 h of quality-selected
live time in 2006-08, 
largely during the commissioning of the instrument \cite{1ES0806_paper}.
A point-like excess (VER\,J0809+523) of 245 VHE $\gamma$-rays is detected,
corresponding to a statistical significance of 6.3$\sigma$.
The VHE spectrum observed between $\sim$300 GeV and
$\sim$700 GeV is soft, with photon index $\Gamma = 3.6 \pm 1.0$, and
the measured flux above 300 GeV is 1.8\% Crab.
No VHE flux variability is observed within limited
statistics.  An SED was generated 
using contemporaneous MWL
observations and is reasonably described by a
one-zone synchrotron-self-Compton (SSC) model.

{\bf W\,Comae} is a bright, nearby ($z = 0.102$), EGRET-detected IBL.  
It was observed with VERITAS for $\sim$40 h of good-quality live time between
January and April 2008 \cite{WCom_paper1}.
A weak detection ($\sim$5$\sigma$) of VHE $\gamma$-ray 
emission (VER\,J1221+282) is found in the overall data set, with 70\% of
the excess (275 $\gamma$-rays, 8.6$\sigma$) 
occurring during a 4-night flare in March 2008.
During the brightest two nights, the VHE flux is
9\% Crab above 200 GeV, and the
VHE spectrum is characterized by a soft 
power law ($\Gamma = 3.8 \pm 0.4$).
While the contemporaneous MWL SED can be described by an
SSC model, an SSC model with an additional
external-Compton (EC) component yields a more
natural set of fit parameters and is therefore preferred.
W\,Comae is the first IBL detected at VHE energies,
and the improvement in the fit from adding an EC component,
not needed in typical HBL modeling, suggests
there may be different underlying processes generating
the VHE $\gamma$-ray emission in different parts of the blazar sequence.
In June 2008, a second VHE flare, $\sim$3 times as bright as the
first, was observed from W\,Comae \cite{WCom_paper2}.  
Modeling of the SED observed during this episode yields 
similar conclusions to the first episode.

{\bf 3C\,66A} is an EGRET-detected IBL with an uncertain 
redshift.  
VERITAS observed
this iconic blazar for 33 h of quality-selected
live time in 2007-08 \cite{3C66A_paper1},
resulting in the discovery of an excess
(VER\,J0222+430) of 1791 VHE $\gamma$-rays (21.2$\sigma$).
The observed integral
flux above 200 GeV is 6\% Crab and shows
evidence for variability on a timescale of days.
The VHE spectrum is soft ($\Gamma = 4.1 \pm 0.4$),
and does not change (within limited statistics) during the nights
of the observed flare \cite{3C66A_paper2}.  
The nearby (separation $\sim$0.12$^{\circ}$) radio galaxy 3C\,66B,
suggested as the possible origin of the MAGIC excess observed from
the same region \cite{MAGIC_3C66} in 2007, is excluded (at a level of
4.3$\sigma$) as the location of the VERITAS excess.
During the flaring episode in October 2008, a simultaneous MWL SED was built
with VERITAS, Fermi-LAT, F-GAMMA, GASP-WEBT, PAIRITEL, MDM, ATOM, 
Swift, and Chandra observations.  While the observed SED can be successfully
modeled using a standard SSC model, with or without an EC component, 
the observed MWL variability can only be explained with the 
addition of the EC component \cite{3C66A_paper2}.

{\bf RGB\,J0710+591} is an HBL at $z = 0.125$ with its
synchrotron peak at an unusually
high-frequency ($10^{19.2}$ Hz).  It
was observed with VERITAS for $\sim$22 h of good-quality live time between
December 2008 and March 2009 \cite{RGBJ0710_paper}. VHE $\gamma$-ray emission 
was discovered from this blazar (5.5$\sigma$) during this
exposure.  The observed VHE flux from VER\,J0710+591 shows no variability
(within statistics) and is 3\% Crab above 300 GeV.  The measured VHE
spectrum between $\sim$310 GeV and $\sim$4.6 TeV is
among the hardest ($\Gamma = 2.69 \pm 0.26$) observed
by VERITAS from a blazar.  Following
the VERITAS detection, the Fermi-LAT collaboration performed
a specific high-energy analysis of their data and found
$\gamma$-ray emission with a very hard spectrum ($\Gamma_{LAT} \sim 1.5$)
between 100 MeV and 300 GeV.  The contemporaneous MWL SED, including
both Fermi-LAT and Swift UVOT/XRT data, is well-described
by an SSC model.  An additional EC component does not improve the fit.
Here it is interesting to note that the inverse-Compton peak of
the SED may be located beyond the highest energy point of
the VERITAS spectrum.

{\bf PKS\,1424+240} is an IBL with an unknown redshift.
It was observed by VERITAS for 28.5 h of quality-selected
live time, between February and June 2009,
because of its inclusion in the Fermi Bright Source List \cite{Fermi_LBAS}.
Soft-spectrum ($\Gamma = 3.8 \pm 0.5$) emission 
above 140 GeV is detected (370 $\gamma$, 8.5$\sigma$) 
with VERITAS \cite{PKS1424_paper}, making this
the first VHE discovery motivated by Fermi-LAT data.  The VHE flux
from VER\,J1427+238,
$\sim$5\% Crab, is steady within the 
observation period, as is the Fermi-LAT flux.  
A contemporaneous MWL SED is well-described
by a one-zone SSC model regardless of
the assumed redshift, provided that $z < 0.66$.  
This limit is determined using a recent
EBL model and combining the Fermi-LAT ($\Gamma_{LAT} = 1.73$)
and VERITAS spectra.
Addition of an EC component does not improve the SED modeling, 
in contrast to the trend seen in other VERITAS IBL detections.

 \begin{figure*}[t]
   \centerline{\includegraphics[width=2.75in]{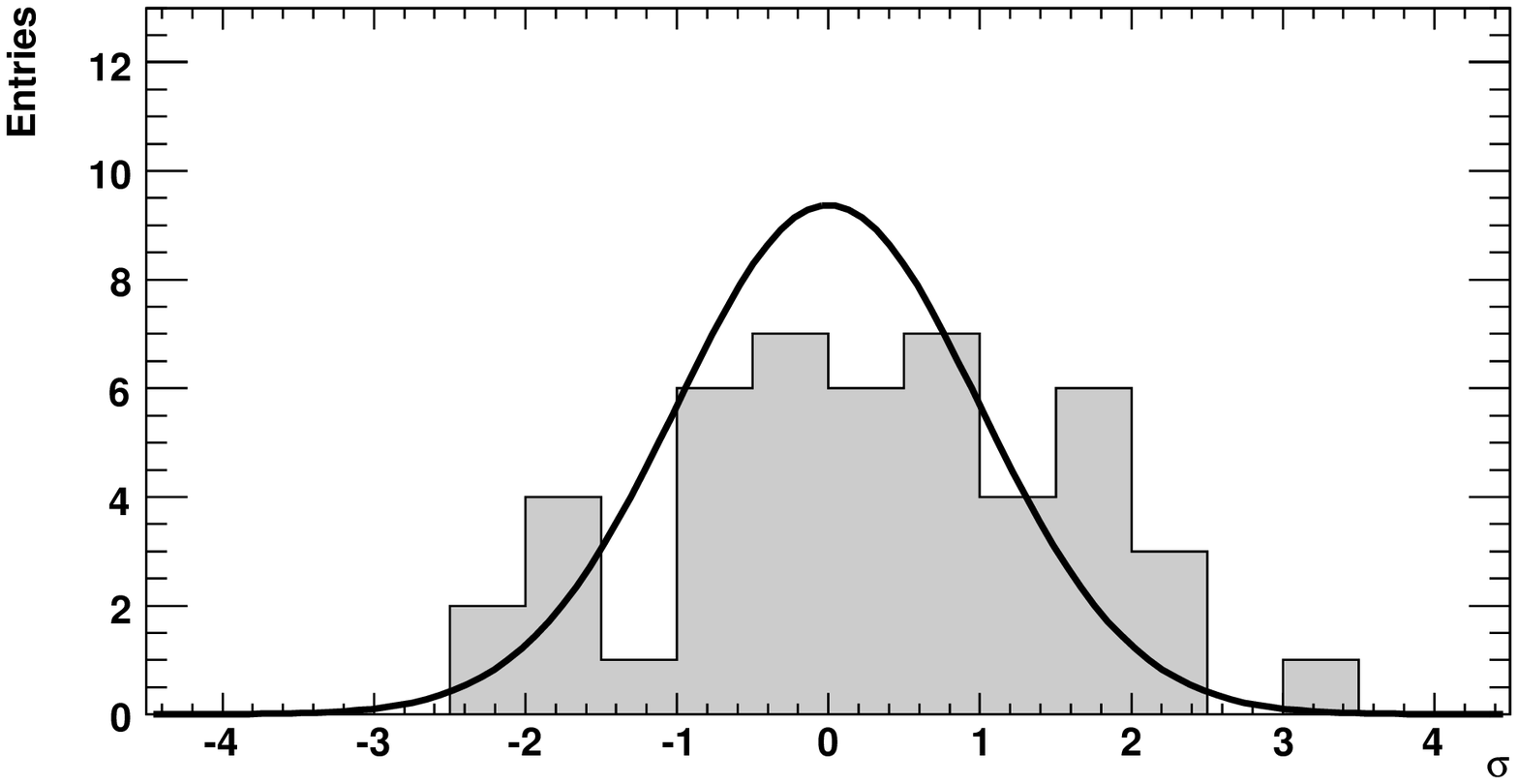}
              \hfil
               \includegraphics[width=2.75in]{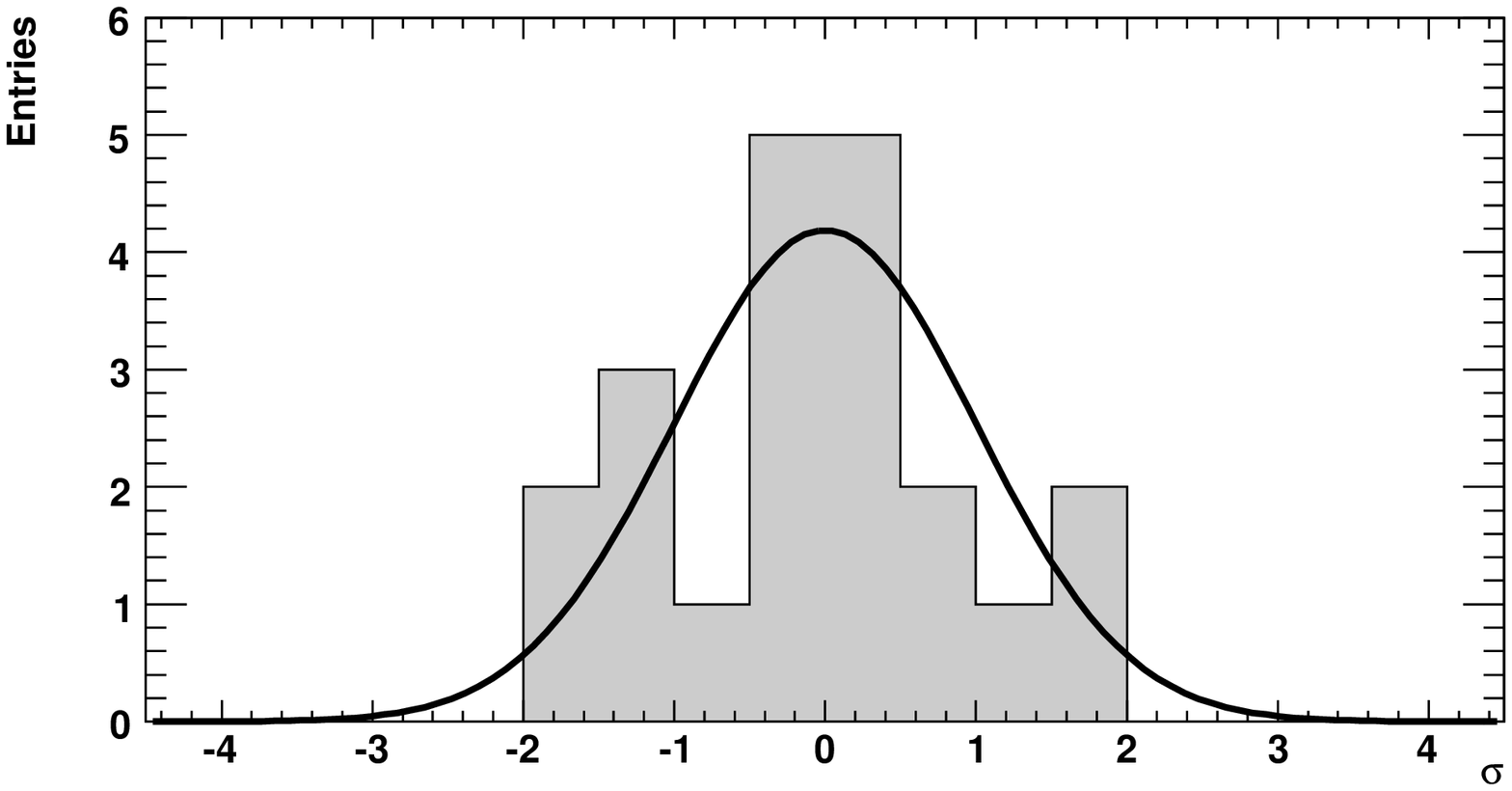}
              }
   \caption{\footnotesize Left: The distribution of significance (soft cuts) observed from 
47 candidate VHE blazars, selected for VERITAS discovery observations
prior to the Fermi launch, during exposures taken 
between 2007 and 2011.  Right:  The distribution of significance (soft cuts) 
observed from  21 candidate VHE blazars, selected using information
in the 1FGL catalog,  during exposures taken in 2009-11.}
   \label{sigma_dists}
 \end{figure*}

\vspace{-0.3cm}
\section{Post-ICRC-2009 Discoveries}
\vspace{-0.2cm}
Nine additional blazars have been detected by VERITAS
since the 2009 ICRC.  Five of these were discovered
as VHE emitters, and for each of these, 
results from a contemporaneous MWL observation campaign,
including VERITAS, Fermi-LAT, Swift XRT/UVOT, and optical data,
are in preparation.  It should also be noted that 
marginal signals ($\sim$4$\sigma$) existed in large 
amounts of VERITAS data ($>$25 h)
for 1ES\,0414+009 and B2\,1215+30 at the time of their announced 
VHE discoveries by the HESS and MAGIC collaborations, respectively.
These two blazars are now strongly detected by VERITAS \cite{Benbow_highlight}.

{\bf RGB\,J0521.8+2112} was initially observed by VERITAS
because of the identification of a nearby
cluster of high-energy photons in 
a simple binned search ($>$30 GeV) of the initial public release
of the Fermi-LAT photon data.  VERITAS observed the likely blazar for
$\sim$15 h of quality-selected live time between October
2009 and January 2010.  An excess of VHE $\gamma$-rays (VER\,J0521+211),
corresponding to a statistical significance of $\sim$16$\sigma$,
is observed from the direction of RGB\,J0521.8+2112.
The observed VHE flux is variable, with an average
value of $\sim$5\% Crab above 200 GeV, and the time-averaged
VERITAS spectrum is soft ($\Gamma = 3.47 \pm 0.19$). 
Follow-up optical spectroscopy of the VERITAS source,
with MDM and the MMT, reveal a continuum 
dominated spectrum typical for BL Lac
objects.  Unfortunately no absorption lines 
are identified and hence the redshift cannot determined.

{\bf RBS\,0413} is an HBL with a redshift of $z = 0.190$. This blazar
was initially selected for observation with VERITAS because it
is one of the X-ray-brightest HBLs in the Sedentary Survey \cite{Sedentary},
and met the selection criteria used to indentify likely VHE emitters in 
\cite{Costamante}.  Deeper observations were later motivated
by the identification of this blazar as
a bright, very hard-spectrum ($\Gamma \sim 1.5$; \cite{Fermi_1FGL}) MeV-GeV
source by the Fermi-LAT collaboration.
RBS\,0413 was observed by VERITAS 
for $\sim$26 h of quality-selected live time 
between September 2008 and January 2010.
An excess of 108 VHE $\gamma$-rays (VER\,J0319+187),
corresponding to a statistical significance of 5.5$\sigma$,
is observed from the direction of RBS\,0413.
The observed VHE flux has an average
value of $\sim$2\% Crab above 200 GeV and does not vary
within the limited statistics. 
The time-averaged VERITAS photon index is $\Gamma = 3.2 \pm 0.7$. 

{\bf 1ES\,0502+675} observations were motivated by the flux and spectrum 
reported in the Fermi-LAT Bright Source List for 
0FGL J0507.9+6739 \cite{Fermi_LBAS, Fermi_1FGL}.  This HBL
was observed by VERITAS for $\sim$30 h of 
quality-selected live time between September 2009
and January 2010. An excess of VHE $\gamma$-rays (VER\,J0507+676),
corresponding to a statistical significance of $\sim$11$\sigma$,
is observed from the direction of the blazar. 
The VHE flux is constant within the observed statistics, with an
average value of $\sim$6\% Crab above 300 GeV, and the time-averaged
VERITAS photon index is $\Gamma = 3.92 \pm 0.35$. 
Although the catalog redshift of this blazar is 0.341,
possibly explaining the softness of the observed spectrum,
the value is from a private communication and 
the original MMT spectrum could not be located.  
Following the VERITAS detection, a $\sim$10 times
more sensitive observation was performed with the
same spectrograph at the MMT.  No significant absorption/emission lines
were observed and hence the redshift is uncertain.

{\bf RX\,J0648.7+1516} observations were motivated
by the identification of a nearby cluster of $>$10 GeV
photons by the Fermi-LAT collaboration.  This unidentified
radio and X-ray emitter was observed by VERITAS 
for $\sim$19 h of quality-selected live time 
between March 4 and April 15, 2010.
An excess of VHE $\gamma$-rays (VER\,J0648+152),
corresponding to a statistical significance of 5.3$\sigma$,
is observed from the direction of RX\,J0648.7+1516.
No strong 
variations are observed in the VHE flux ($\sim$2\% Crab above 300 GeV),
and the time-averaged VERITAS spectrum is very soft ($\Gamma = 4.4 \pm 0.8$). 
Follow-up optical spectroscopy was performed 
with the Shane 3-m telescope at Lick Observatory. A 
continuum dominated spectrum typical of BL Lac objects is observed,
along with weak absorption lines compatible with $z = 0.179$.

{\bf 1ES\,1440+122} is an IBL (borderline HBL), 
with a well-determined redshift of $z = 0.162$.
This hard-spectrum Fermi source ($\Gamma \sim 1.8$; \cite{Fermi_1FGL}) 
was selected for observation with VERITAS because it
was recommended as a likely VHE emitter by \cite{Costamante}.  
The blazar was observed by VERITAS 
for $\sim$47 h of quality-selected live time 
between May 2008 and June 2010.
An excess of VHE $\gamma$-rays (VER\,J1442+120),
corresponding to a statistical significance of 5.5$\sigma$,
is observed from the direction of 1ES\,1440+122.
The observed VHE flux does not vary within limited
statistics and has an average
value of $\sim$1\% Crab above 200 GeV.  The time-averaged
VERITAS photon index is $\Gamma = 3.4 \pm 0.7$. 

\vspace{-0.3cm}
\section{Non-detected Blazars}
\vspace{-0.2cm}
VERITAS has observed
approximately 90 blazars in search of VHE emission.
Unfortunately, most of these observations did not
successfully identify a new VHE emitter.  The distribution
of significance observed from 47 targets, taken from 
the X-ray/EGRET-selected 
list used between September 2007 and June 2009, 
is shown in Figure~\ref{sigma_dists}.
The distribution is skewed towards positive values
of significance.  Stacking the result from all these observed
blazars ($\sim$280 h exposure) 
yields an excess of 656 events (4.1$\sigma$) using selection
cuts optimized for soft-spectrum sources\footnote{The excess is 294 events 
(3.4$\sigma$) using the standard event-selection criteria.}.  This excess is not
seen when stacking results from an identical analysis of
other non-detected extragalactic non-blazar observations.  
A similar VERITAS result was presented at the 2009
ICRC \cite{Benbow_ICRC09}.  Interestingly, seven of
the objects contributing to that excess have since been
detected at VHE (five by VERITAS) and are removed from the stacking
here. In addition, several targets now have larger exposures, and
some objects from the initial target list were observed after June 2009
and are now included.  Twenty-one Fermi-selected objects
were observed, but not detected, in the 2009-10 season\footnote{Further data 
were taken in the 2010-11 season.}.  The
distribution of significance observed is shown in Figure~\ref{sigma_dists}.
There is no evidence for any stacked excess ($\sim$154 h exposure).  
This may be due
to the lack of redshift measurements for many of these
objects.  Analysis of addtional Fermi-selected objects observed
during the 2010-11 season is in preparation.

\vspace{-0.3cm}
\section{Conclusion}
\vspace{-0.2cm}
VERITAS has detected 20 blazars, including 10 for the first
time at VHE. Every VERITAS discovery has initiated
a deep MWL observation campaign and corresponding modeling
of the resulting SEDs.  In general, one-zone SSC models are
reasonable descriptions of the data for the population of VHE HBL objects. 
Modeling of the objects in the VHE IBL population usually 
requires an additional EC emission component.
Clearly, a deeper population of VHE blazars is desirable.
In the future, VERITAS will continue to observe the targets in the initial
X-ray/EGRET-selected list, particularly those
with marginal excesses in previous VERITAS
observations, as well as new targets identified
using Fermi-LAT.  The strategy for the Fermi-LAT targets 
may focus on deeper observations of objects,
particularly non-HBLs, with large redshifts ($z > 0.3$),
and on ToO observations of those that are flaring, as identified through 
automatic monitoring by the VERITAS
collaboration of the public Fermi-LAT data.

\vspace{0.2cm}
{\footnotesize
This research is supported by grants from the US Department of Energy, 
the US National Science Foundation, and the Smithsonian Institution, 
by NSERC in Canada, by Science Foundation Ireland, and by STFC in the UK. 
We acknowledge the excellent work of the technical support staff at 
FLWO and the collaborating institutions in the 
construction and operation of the instrument.}

\clearpage

\end{document}